\documentclass[11pt]{article}
\usepackage{spconf,amsmath,epsfig}
\usepackage{graphicx} 
\usepackage{graphics}
\usepackage{upgreek}
\usepackage{amsfonts,amssymb}
\usepackage{sectsty}
\usepackage{makecell}
\usepackage{subfigure}
\usepackage[usenames]{color}
\usepackage{rotating}
\usepackage{verbatim}
\usepackage{framed}
\usepackage{mathtools}
\usepackage{cite}
\usepackage{bbm}
\usepackage{bm}
\usepackage{amsfonts}

\def\bff{{\mathbf{f}}}

\def\bm{{\mathbf{m}}}

\def\bu{{\mathbf{u}}}

\def\bx{{\mathbf{x}}}
\def\by{{\mathbf{y}}}
\def\bz{{\mathbf{z}}}
\def\b0{{\mathbf{0}}}

\def\bI{{\mathbf{I}}}

\def\bK{{\mathbf{K}}}

\def\bQ{{\mathbf{Q}}}

\def\bS{{\mathbf{S}}}

\def\bX{{\mathbf{X}}}

\def\bZ{{\mathbf{Z}}}

\def\btheta{{\boldsymbol{\theta}}}

\def\bmu{{\boldsymbol{\mu}}}

\def\bSigma{{\boldsymbol{\Sigma}}}

\def\rd{{\mathrm{d}}}

\newcommand{\p}{{\mathrm p}}
\newcommand{\q}{{\mathrm q}}

\def\R{{\mathbb{R}}}

\usepackage{hyperref} 
\hypersetup{
    bookmarks=true,         
    unicode=false,          
    pdftoolbar=true,        
    pdfmenubar=true,        
    pdffitwindow=false,     
    pdfstartview={FitH},    
    pdftitle={DGP},    
    pdfauthor={NAMES, NAMES, NAMES},     
    pdfsubject={DGP},   
    pdfcreator={NAMES, NAMES, NAMES},   
    pdfproducer={DGP}, 
    pdfkeywords={Forward}{inverse}{modelling}{Gaussian process}{kernels}, 
    pdfnewwindow=true,      
    colorlinks=true,       
    linkcolor={blue},          
    citecolor=blue,        
    filecolor=blue,      
    urlcolor=blue           
}

\title{Deep Gaussian Processes for geophysical parameter retrieval}
\name{Daniel Heestermans Svendsen$^{1,*}$, Pablo Morales-\'Alvarez$^{2,*}$, Rafael Molina$^2$, Gustau Camps-Valls$^1$\thanks{{\bf Preprint, Paper published in IGARSS 2018 - 2018 IEEE International Geoscience and Remote Sensing Symposium, Valencia, 2018, pp. 6175-6178, doi: 10.1109/IGARSS.2018.8517647.}}
\thanks{Research funded by the European Research Council (ERC) under the ERC-CoG-2014 SEDAL project (grant agreement 647423), the Spanish Ministry of Economy and Competitiveness through projects TIN2015-64210-R, DPI2016-77869-C2-2-R, and the
Spanish Excellence Network TEC2016-81900-REDT. PMA received financial support through \emph{La Caixa} Fellowship for Doctoral Studies (\emph{La Caixa} Banking Foundation, Barcelona, Spain).}\thanks{* The first two authors contributed equally.}}
\address{$^1$Image Processing Laboratory (IPL), Universitat de Val\`encia, Spain\\
$^2$Computer Science and Artificial Intelligence Department, Universidad de Granada, Spain}

\begin{document}
%
\maketitle
\begin{abstract}
This paper introduces deep Gaussian processes (DGPs) for geo-physical parameter retrieval. Unlike the standard full GP model, the DGP accounts for complicated (modular, hierarchical) processes, provides a efficient solution that scales well to large datasets, and improves prediction accuracy over standard full and sparse GP models. We give empirical evidence of performance for estimation of surface dew point temperature from infrared sounding data. 
\end{abstract}
%
\section{Introduction} \label{sec:int}

Remote sensing applications often involve solving a complex model inversion problem to estimate bio-geo-physical parameters from observations. The problem is very challenging as it involves nonlinear input-output relations, as well as a wide diversity of noise sources and distortions. In addition, very often, the goal is to invert {\em metamodels}, that is, combinations of submodels that are coupled together. Radiative transfer models (RTM) describe the processes occurring at different scales (e.g. at leaf, canopy and atmospheric levels) with different complexities. The overall process is thus complicated and with different amounts of uncertainties that accumulate and propagate. 

Inverting such models is complicated and several strategies exist: interpolation on look-up tables, direct physically-based inversion, or statistical approaches. Physics-based retrieval is often very costly and needs a broad knowledge about the governing mechanisms. In recent years, the remote sensing community has turned to more efficient and agnostic statistical {\em hybrid} approaches for model inversion~\cite{CampsValls11mc}. After generating pairs of inputs (state vectors) and outputs (radiances) by running an RTM code, the problem boils down to performing inverse modeling via standard regression. Approximating arbitrary nonlinear functions from data is a solid field of machine learning where many successful methods are available.

In the last few years, data-driven statistical learning algorithms have attained outstanding results in the estimation of climate variables and related geo-physical parameters at local and global scales~\cite{CampsValls09wiley,CampsValls11mc}. These algorithms avoid complicated assumptions and provide flexible non-parametric models that fit the observations using large heterogeneous data. 
In particular, during the last decade, the use of Gaussian Processes (GPs) in model inversion and emulation of physical models in the geosciences has gained popularity in general~\cite{campsvalls16grsm}, and for atmospheric parameter retrieval in particular~\cite{Camps-Valls20121759}. However, GPs have two important shortcomings: their high computational cost ($n^3$ in training and $n^2$ in testing), and the limited expressiveness of standard kernel functions (often restricted to the exponentiated quadratic family). On top of these, one should note the hierarchical modularity of the physical models to invert. This actually discourages the use of a plain, shallow GP model to account for all nonlinear feature relations.  

To address the aforementioned problems, in this paper, we explore deep GP (DGP) regression for geophysical parameter retrieval and model inversion. DGPs were originally introduced in~\cite{damianou2013deep} and further analyzed in~\cite{damianou_deep_2015}.  
We will resort to the very recent DGP inference procedure in~\cite{salimbeni_doubly_2017}, which has provided excellent results in a wide range of regression problems.  
We will show that DGPs overcome the limitations of standard GPs,
in particular how they: 1) account for complicated (modular, hierarchical) processes encoded in the RTM, 2) return an efficient solution that scales well to large datasets, and 3) improve prediction accuracy over standard full and sparse GP models. We illustrate performance in prediction of surface temperature and moisture  
from superspectral infrared sounding data~\cite{aires02}.  
 
\section{Probabilistic Model and Inference} \label{sec:theo}

Gaussian Processes \cite{Rasmussen2006} are non-parametric probabilistic state-of-the-art models for functions, that are successfully used in supervised and unsupervised learning.
Notationally, for input-output data $\{(\bx_i,y_i)\in\R^d\times\R\}_{i=1}^n$, a GP models the underlying dependence with latent variables $\{f_i = f(\bx_i)\in\R\}_{i=1}^n$ that jointly follow a Gaussian $\mathcal{N}\left(\bff|\mathbf{0},\bK\right)$.
The kernel matrix $\bK = (k(\bx_i,\bx_j))_{i,j}$ depends on the kernel function $k$, which encodes the properties (e.g. smoothness) of the modeled functions. In regression problems, the observation model of the outputs $y_i$ given the latent variables $f_i$ is usually defined by the Gaussian $\p(y_i|f_i,\sigma^2)=\mathcal{N}(y_i|f_i,\sigma^2)$.

This likelihood allows us to integrate out $\bff$ and compute the posterior $\p(\bff|\by)$ in closed form solution (parameters are omitted for simplicity) \cite{Rasmussen2006}. However, this requires inverting the $n\times n$ matrix $(\bK+\sigma^2 \bI)$, which scales cubically, $\mathcal{O}(n^3)$. This constraint makes GP prohibitive for large scale applications. Here, sparse GP approximations become the more efficient pathway.

Additionally, GPs are limited by the expressiveness of the kernel function. Although very complex kernels can be tailored to the data \cite{Rasmussen2006}, in practice this may become unfeasible, as it requires a thorough application-specific knowledge and usually comes with a large amount of hyperparameters to estimate, which may cause overfitting. Alternatively, DGPs allow for modelling very complex data through a hierarchy of GPs that only use simple kernels with few hyperparameters.
\subsection{Sparse GP approximations}\label{sec:sparseGP}

The most popular approach to dealing with the computational burden of GPs is to introduce $m\ll n$ \emph{inducing points} $\bu=(u_1,\dots,u_m)$ which the inference is based on. These are GP realizations at the \emph{inducing locations} $\bZ=\{\bz_1,\dots,\bz_m\}\subset\R^d$, just like $\bff$ is at the inputs $\bX=\{\bx_1,\dots,\bx_n\}$ \cite{bauer2016}.
One of the most widely used methods is the \emph{Fully Independent Training Conditional (FITC)} introduced in \cite{snelson2006sparse}.

FITC approximates the GP prior model\footnote{Among the sparse GP approximations based on inducing points, two families are distinguished: those that approximate the GP prior and perform exact inference (where FITC is the most popular), and those that utilize the exact prior but perform approximate inference (VFE \cite{titsias2009} is the main example). Both families are compared in detail in \cite{bauer2016}. A sparse GP approximation without inducing points was recently applied to remote sensing data \cite{morales2017}.}
by assuming: i) conditional independence between train and test latent variables $\bff,\bff_*$ given the inducing points $\bu$; and ii) a factorized (fully independent) distribution for $\bff$ given $\bu$. The effective GP joint prior for FITC (which replaces the exact $\p(\bff,\bff_*)$) is:
\begin{equation}
    \q(\bff,\bff_*) = \mathcal{N}\left(\mathbf{0},\left[
    \begin{array}{cc}
       \bQ_{ff} + \mathrm{diag}(\bK_{ff}-\bQ_{ff})  & \bQ_{f*} \\
        \bQ_{*f} & \bK_{**} 
    \end{array}
    \right]
    \right),
\end{equation}
where we abbreviate $\bQ_{ab}=\bK_{au}\bK_{uu}^{-1}\bK_{ub}$. With this approximation, the observation model $\p(\by|\bff,\sigma^2)$ can be integrated out and the new matrix to be inverted is $(\bQ_{ff} + \mathrm{diag}(\bK_{ff}-\bQ_{ff})+\sigma^2\bI)$. Applying the Woodbury matrix identity, this low-rank-plus-diagonal matrix can be inverted with $\mathcal{O}(nm^2)$ complexity. Notice that this is linear in the training set size $n$. 

\subsection{Deep Gaussian Processes} \label{sec:DGP}

In standard (single-layer) GPs, the output of the GP is directly used to model the observed response $\by$. However, this output could be used to define the input locations of another GP. If this is repeated $L$ times, we obtain a hierarchy of GPs that is known as a Deep Gaussian Process (DGP) with $L$ layers. This is analogous to the structure of deep neural networks, which are a cascade of generalized linear models \cite[Chapter 6]{damianou_deep_2015}.

DGPs were first introduced in \cite{damianou2013deep}, where the authors performed approximate variational inference analytically.
However, in each layer they use a set of latent variables that ends up inducing \emph{independence across layers} in the posterior distribution. This uncorrelated posterior fails to express the complexity of the deep model, and is not realistic in practice. To overcome this problem, we follow the recent inference procedure in \cite{salimbeni_doubly_2017}, which keeps a strong conditional independence in the posterior by marginalizing out the aforementioned set of latent variables. In exchange, analytic tractability is sacrificed. However, we will see that the structure of the posterior allows one to efficiently sample from it and use Monte Carlo approximations. As will be justified later, this approach is called \emph{Doubly Stocashtic Variational Inference} \cite{salimbeni_doubly_2017}.

\begin{figure}
    \centering
    \includegraphics[width=7cm]{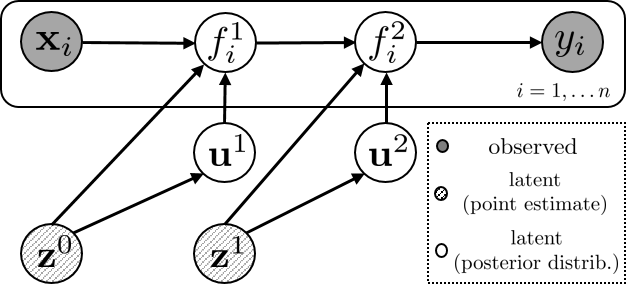}
    \caption{Graphical representation of the DGP model used here.}
    \label{fig:model}
\end{figure}

DGPs can be used for regression by placing a Gaussian likelihood after the last layer. For notation simplicity, in this work the dimensions of the hidden layers will be fixed to one\footnote{This can be generalized straightforwardly, see both approaches \cite{damianou2013deep, salimbeni_doubly_2017}.}. 
However, exact inference in DGP is intractable (not only computationally expensive as in GPs), as it involves integrating out latent variables that are used as inputs in the next layer (i.e. they appear inside a complex kernel matrix). To overcome this, $m$ inducing points $\bu^l$ at inducing locations $\bz^{l-1}$ are introduced at each layer $l$. Interestingly, we will see that this sparse formulation also makes DGP scale well to large datasets. 
For observed $\{\bX,\by\}$, the regression joint model is
\begin{equation}\label{eq:jointModel}
    \p(\by,\{\bff^l,\bu^l\}_{1}^L)=\p(\by|\bff^L)\prod_{l=1}^L \p(\bff^l|\bu^l;\bff^{l-1},\bz^{l-1})\p(\bu^l;\bz^{l-1}).
\end{equation}
Here, $\bff^0=\bX$, and each factor in the product is the joint distribution over $(\bff^l,\bu^l)$ of a GP in the inputs $(\bff^{l-1},\bz^{l-1})$, but rewritten with the conditional probability given $\bu^l$\footnote{Notice that a semicolon is used to specify the inputs of the GP.}.
Figure \ref{fig:model} shows a graphical representation of the described model.

\subsubsection{Doubly Stochastic Variational Inference}\label{sec:inference}

The approach followed in \cite{salimbeni_doubly_2017} uses variational inference, where a lower bound of the marginal log-likelihood is maximized to find the optimal posterior among a family of distributions. The proposed posterior is
\begin{equation}\label{eq:jointPosterior}
    \q(\{\bff^l,\bu^l\}_{l=1}^L)=\prod_{l=1}^L \p(\bff^l|\bu^l;\bff^{l-1},\bz^{l-1})\q(\bu^l).
\end{equation}
The first factor is the prior conditional of eq.~\eqref{eq:jointModel} and keeps correlations between layers. The second is taken Gaussian with mean $\bm^l$ and full covariance $\bS^l$ (variational parameters to be estimated). With this posterior, the next \emph{evidence lower bound (ELBO)} for the marginal log-likelihood is obtained\footnote{The prior conditionals of eq.~\eqref{eq:jointPosterior} smartly cancel with those of eq.~\eqref{eq:jointModel}.}:
\begin{multline}\label{eq:lowerBound}
    \log\p(\by)=\log\int\frac{\q(\{\bff^l,\bu^l\})}{\q(\{\bff^l,\bu^l\})}\p(\by,\{\bff^l,\bu^l\})\rd\bff^l\rd\bu^l\geq \\
    \sum_{i=1}^n\mathbb{E}_{\q(f^L_i)}[\log\p(y_i|f^L_i)] - \sum_{l=1}^L\mathrm{KL}(\q(\bu^l)||\p(\bu^l;\bz^{l-1})).
\end{multline}
The second term is tractable, as the KL divergence between Gaussians is known \cite{Rasmussen2006}. However, the expectation involves the marginals of the posterior at the last layer, $\q(f^L_i)$. Next we see that, whereas this distribution is analytically intractable, it can be sampled efficiently using univariate Gaussians.

Marginalizing out the inducing points in eq.~\eqref{eq:jointPosterior}, the posterior for the GP layers $\{\bff^l\}$ is
\begin{equation}\label{eq:posteriorF}
    \q(\{\bff^l\})\!=\!\prod_{l=1}^L \q(\bff^l|\bm^l,\bS^l;\bff^{l-1},\bz^{l-1})\!=\!\prod_{l=1}^L \mathcal{N}(\bff^l|\tilde\bmu^l,\tilde\bSigma^l),
\end{equation}
where the vector $\tilde\bmu^l$ is given by $[\tilde\bmu^l]_i=\mu_{\bm^l,\bz^{l-1}}(f^{l-1}_i)$ and the $n\times n$ matrix $\tilde\bSigma^l$ by $[\tilde\bSigma^l]_{ij}=\Sigma_{\bS^l,\bz^{l-1}}(f^{l-1}_i,f^{l-1}_j)$. The specific form of the functions $\mu_{\bm^l,\bz^{l-1}}$ and $\Sigma_{\bS^l,\bz^{l-1}}$ can be found in \cite[Eqs. (7-8)]{salimbeni_doubly_2017}. The key point here is to observe that, although the distribution in eq.~\eqref{eq:posteriorF} is fully coupled between layers (and thus the posterior in the last layer is analytically intractable), the $i$-th marginal at each layer $\mathcal{N}(f^l_i|[\tilde\bmu^l]_i,[\tilde\bSigma^l]_{ii})$ only depends on the corresponding $i$-th input of the previous layer. This allows one to recursively sample $\hat f_i^1\to \hat f_i^2\to\dots\to \hat f_i^L$ from all the layers up to the last one by means of univariate Gaussians. Specifically, $\varepsilon^l_i\sim\mathcal{N}(0,1)$ is first sampled and then for $l=1,\dots,L$:
\begin{equation}\label{eq:sample}
    \hat f_i^l = \mu_{\bm^l,\bz^{l-1}}(\hat f_i^{l-1})+\varepsilon^l_i\cdot\sqrt{\Sigma_{\bS^l,\bz^{l-1}}(\hat f_i^{l-1},\hat f_i^{l-1})}.
\end{equation}
Now, the expectation in the ELBO (recall eq.~\eqref{eq:lowerBound}) can be approximated with a Monte Carlo sample generated with eq.~\eqref{eq:sample}. This provides the first source of stochasticity. Since the ELBO factorizes across data points and the samples can be drawn independently for each point $i$, scalability is achieved through sub-sampling the data in mini-batches. This is the second source of stochasticity. The ELBO is maximized w.r.t. the variational parameters $\bm^l,\bS^l$, the inducing locations $\bz^l$, and the kernel hyperparameters $\btheta^l$ (which, to alleviate the notation, have not been included in the equations). 
The complexity to evaluate the ELBO and its gradients is $\mathcal{O}(nm^2L)$. The code is integrated within GPflow (a GP framework built on top of Tensorflow) and is publicly available\footnote{https://github.com/ICL-SML/Doubly-Stochastic-DGP}.

To predict in a new $x_*$, eq.~\eqref{eq:sample} is used to sample $S$ times\footnote{Results become stable after a few samples. Here, $S$ was set to 200.} from the posterior up to the $(L-1)$-th layer using the test location as initial input. This yields a set $\{f_*^{L-1}(s)\}_{s=1}^S$ with $S$ samples. Then, the density over $f_*^L$ is given by the Gaussian mixture (recall that all the terms in eq.~\eqref{eq:posteriorF} are Gaussians):
\begin{equation*}
    \q(f_*^L)=\frac{1}{S}\sum_{s=1}^S \q(f_*^L|\bm^L,\bS^L;f_*^{L-1}(s),\bz^{L-1}).
\end{equation*}

\section{Experimental results } \label{sec:exp}

We compare prediction error and training times with a standard full GP and FITC, the most widely used sparse GP approximation. 

\subsection{Data description}

We focus on assessing the performance of the DGP for predicting relative humidity at surface level, measured as dew point temperature $T_{dew}$ [K], using data from the Infrared Atmospheric Sounding Interferometer (IASI). The IASI data are point measurements of approximately 25 km diameter with 8461 spectral components, ranging in the infrared emission spectra from 645 to 2760 cm$^{-1}$ with 0.25 cm$^{-1}$ resolution. 
We used the simulated database in \cite{chevallier2002sampled}, which consists of 13456 datapoints generated using the optimal spectral sampling (OSS) infrared radiative transfer code \cite{moncet2008infrared}. 
First,  we  remove  certain  bands  from  the  spectrum  that do not contain useful information for retrieval, reducing the data to 4699 spectral components, following the same scheme proposed in \cite{Camps-Valls20121759}. Dimensionality reduction is then performed with  principal component analysis (PCA), projecting the spectra down to a 13 dimensional subspace, where 99.96\% of the data variance is retained. 

\subsection{Predicting dew point temperature}

\begin{figure}[htb]
\centerline{
\includegraphics[width=7cm]{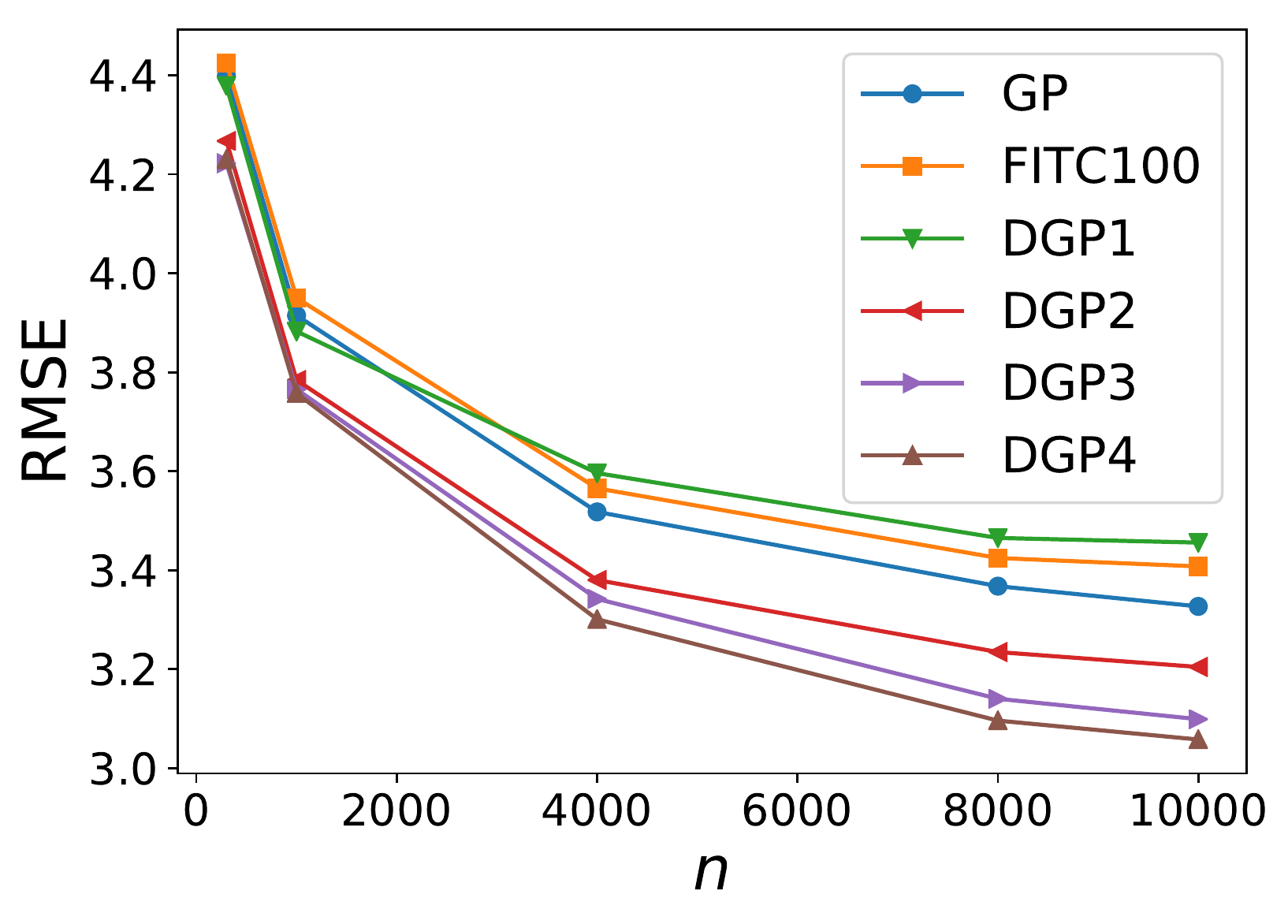}}
\vspace{-0.4cm}
\caption{Performance of the compared methods as a function of the training set size. DGP2-4 outperform the full GP, which in turn is better than its sparse approximations DGP1 and FITC.}
\label{fig:rmse} 
\end{figure}

The spectra were randomly split into training sets of sizes 300, 1000, 4000, 8000 and 10000, and the rest are left for testing. The resulting root mean squared error (RMSE) is shown in Fig. \ref{fig:rmse}, which is averaged over several runs. We compare the following GP-based models:\\

\noindent \textbf{GP:} The standard full GP, using an exponentiated quadratic (EQ) kernel function. The computational cost of training scales like $\mathcal{O}(n^3)$ with the number of training points.

\noindent \textbf{FITC:} Introduced in Section \ref{sec:sparseGP}, it is among the most popular sparse GP approximations. The EQ kernel is used, and the code is taken from GPflow \url{https://github.com/GPflow}. The cost of training scales like $\mathcal{O}(nm^2)$, and the number inducing points is set to $m=100$.

\noindent \textbf{DGP1-4:} DGP described in Section \ref{sec:DGP}
with 1-4 layers and $m=100$ inducing inputs per layer. The number of hidden units per layer is equal to the dimension of the input data (13). Note that DGP1 is just a sparse GP using the inference scheme of Sec. \ref{sec:inference}. The computational cost is $\mathcal{O}(nm^2L)$.\\

As theoretically expected, the full GP consistently outperforms its sparse approximations FITC and DGP1. The DGP2-4 take advantage of their hierarchical structure and achieve even lower RMSE than the full GP. Moreover, as the amount of training data increases, the gap between DGP2 and DGP3,4 widens, indicating that there is a complex structure in the data which can be learned when combining a sufficiently complex model (3 or more layers) with sufficient amounts of data.

Interestingly, we see that, for a small amount of training data ($n=300$), the 2-,3- and 4-layer DGPs, do not overfit and perform better than the single-layer sparse models and the full GP. This is in spite of having many parameters to fit compared to the amount of data, showing that the Bayesian training scheme adequately penalizes complexity.

\begin{figure}[htb]
\centerline{
\includegraphics[width=7cm]{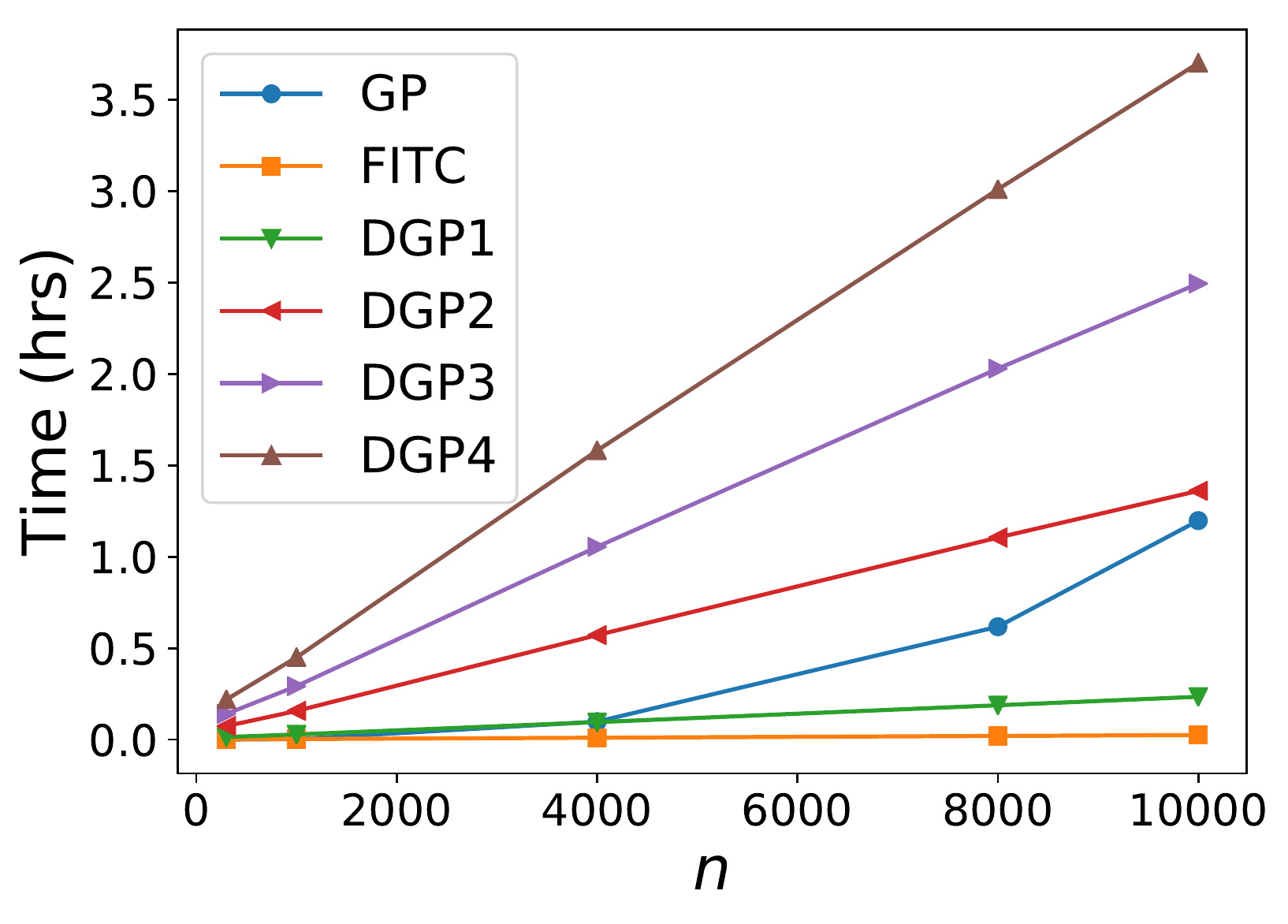} }
\vspace{-0.4cm}
\caption{Training time of the compared methods for increasing $n$. We see how the DGP and FITC scale linearly with $n$ while the full GP scales cubically, making it unfit for big data.}
\label{fig:time} 
\end{figure}

Fig.~\ref{fig:time} shows the training times for the different models. As theoretically expected, all models scale linearly except for the GP, which is $\mathcal{O}(n^3)$. This makes it impossible to train a GP on a dataset of magnitude $n \sim 10^5$ \cite{Rasmussen2006}. However, DGPs can be trained within a reasonable time span on big datasets. Thus, they constitute a very powerful alternative to classical sparse GP approximations (such as FITC), especially for datasets with a complex underlying structure.

\section{Conclusions} \label{sec:con}

We introduced deep GPs and the doubly stochastic variational inference procedure for remote sensing applications. We showed how DGP benefits from its hierarchical structure and consistently outperforms a full GP in superspectral infrared sounding data. Moreover, unlike the full GP, its sparse formulation makes it scale linearly with the training set size. Thus, DGP is further compared against FITC, the most widely used classical sparse GP approximation. The experimental results suggest that DGP performs better than current state-of-the-art GP-based methods in large scale remote sensing datasets, especially when there is a complex underlying process generating the data.

\end{document}